\documentclass{article}[11pt]
\def\journal#1#2#3#4{ {#1} {\bf #2}, {#3}\  ({#4})}

\def\ibid{\journal{\em ibid.}}

\def\MPLA{\journal{Mod.\ Phys \ Lett. \ {\bf A}}}

\def\NPProc{\journal{Nucl.\ Phys.\ Proc.\ Suppl}}

\def\NPB{\journal{Nucl.\ Phys.\ {\bf B}}}
\def\NuovoCimento{\journal{Nuovo Cimento}}

\def\PLA{\journal{Phys.\ Lett.\ {\bf A}}}
\def\PLB{\journal{Phys.\ Lett.\ {\bf B}}}
\def\PhysRev{\journal{Phys.\ Rev.}}

\def\PRD{\journal{Phys.\ Rev.\ {\bf D}}}
\def\PRL{\journal{Phys.\ Rev.\ Lett.}}

\begin{document}

\newcommand{\gn}{\mbox{$\gamma_{\stackrel{}{5}}$}}
\newcommand{\adag}{a^{\dagger}}
\newcommand{\aprime}{a^{\prime}}
\newcommand{\aprimedag}{a^{\prime \dagger}}
\newcommand{\AdagqL}{A^{\dagger}_{q,L}}
\newcommand{\AdagprimeqL}{A^{\prime\dagger}_{-q,L}}
\newcommand{\AqL}{A^{}_{q,L}}
\newcommand{\BqR}{B^{}_{-q,R}}
\newcommand{\akL}{a^{}_{k,L}}
\newcommand{\bkR}{b^{}_{-k,R}}
\newcommand{\BdagkR}{B^{\dagger}_{-k,R}}
\newcommand{\adagkL}{a^{\dagger}_{k,L}}
\newcommand{\bdagkR}{b^{\dagger}_{-k,R}}
\newcommand{\chiL}{\chi_{_L}}
\newcommand{\chiR}{\chi_{_R}}
\newcommand{\eps}{\epsilon}
\newcommand{\psibar}{\bar{\psi}}
\newcommand{\psidag}{\psi^{\dagger}}
\newcommand{\psiL}{\psi_{_{L}}}
\newcommand{\psiR}{\psi_{_{R}}}
\newcommand{\thetap}{\theta_{p}}
\newcommand{\costhetap}{\cos{\thetap}}
\newcommand{\sinthetap}{\sin{\thetap}}
\newcommand{\Tprime}{T'}
\newcommand{\Tprimesq}{T^{'2}}
\newcommand{\x}{\vec{x},t}
\newcommand{\xPrime}{\vec{x} - \hat{n} (t - t' ), t'}
\newcommand{\xPrimet}{\vec{x} + \hat{n} (t - t'), t'}
\newcommand{\y}{\vec{y}, y_o}

\vspace*{-.15in}

\hspace*{\fill}\fbox{CCNY-HEP-01-11}

\begin{center}
{\Large {\bf Oscillations of Faster than Light Majorana Neutrinos: \\
A Causal Field Theory\fnsymbol{footnote}\footnote
{\samepage \sl
\noindent \parbox[t]{138mm}{ \noindent
                       This work has been supported in part by a 
                       grant from PSC-BHE of CUNY. 
                           }    

}
}
}\\
\baselineskip 5mm
\ \\
Ngee-Pong Chang (npccc@sci.ccny.cuny.edu)\\
Department of Physics\\
The City College \& The Graduate Center of \\
The City University of New York\\
New York, N.Y. 10031\\
\  \\
May 9, 2001  \\
\end{center}

\vspace*{-.15in}

\noindent\hspace*{\fill}\parbox[t]{5.5in}{
        \hspace*{\fill}{\bf Abstract}\hspace*{\fill} \\
        {\em

	In this paper, we carry out the canonical quantization of the field theory 
	of an interacting tachyonic Majorana neutrino.   
	We show how micro-causality is preserved in the physical scattering
	matrix elements between the in and out vacua. \\

      The phenomenology of this radical proposal is nevertheless compatible with normal
	timelike oscillations.
        } 
                              }\hspace*{\fill} \\             

\begin{flushleft}
\end{flushleft} 


\section{Introduction}
	
	Recent data from Super-Kamiokande\cite{superkamiokande} suggest that 
	neutrinos oscillate and are therefore massive.  But does this mean 
	that they are necessarily Dirac neutrinos, or can they be Majorana 
	neutrinos ?

	There has been much recent interest on this possibility\cite{majorana-recent}. 
	In this paper, we propose an observation that the neutrino 
	can not only be a Majorana particle, but, in an interesting 
	twist, has a tachyonic\cite{tachyon} mass. (See 
	ref\cite{tachyon-chodos} for earlier work on a tachyonic
	neutrino.)  We shall show that the phenomenology of such
	a radical proposal is nevertheless compatible with normal
	timelike oscillations. \\


{\large \bf Problems with Tachyon Field}

	Earlier studies of tachyon field theory encountered many 
	difficulties.
	It is well-known that a tachyon field $\psi_{_{sp}}$ containing 
	only spacelike superluminal modes ($k > m$) does not obey 
	micro-causality.
      The equal time anti-commutator $\{ \psi_{_{sp}} (\vec{x}, 0), 
	\psi^{\dagger}_{_{sp}}( \vec{y}, 0)\}$ 
	is not a simple spatial delta function, but has a space-like 
	tail that destroys micro-causality.

	If in trying to solve the micro-causality problem, you include 
	modes with $k < m$, you encounter complex solutions, and the 
	problem becomes one of physical interpretation of exponential 
	runaway states. \\

{\large \bf Majorana Field Theory} 

	In this paper, we study the quantum field theory of a pair of opposite
	metric massless left-handed fields (see eq.(\ref{eq-Lagrangian})) whose 
	Majorana interactions lead to tachyonic modes for the physical
	matrix elements.  We carry through the complete canonical quantization
	in order to clarify the issues of superluminal propagation and 
	its impact on micro-causality.
	We find that, in addition to the expected Lorentz 
	invariant space-like measure
\begin{equation}
	\int d^4 p \; \delta(p \cdot p - m^2)
\end{equation}
	for the neutrino field, there is a missing Lorentz invariant measure that
	must be included in the field.  It is a measure for complex momenta 
\begin{equation}
	\int d^4 q \; \delta(q \cdot q - m^2) \; \theta(m^2 - q \cdot \bar{q})
\end{equation}
	Here, we are using the metric such that
\begin{eqnarray}
	q \cdot q & = & \vec{q} \cdot \vec{q} \;\;- q_o \cdot q_o \\
	q \cdot \bar{q} & = & \vec{q} \cdot \vec{q}\,{}^{*} - q_o \cdot q_o^{*} 
\end{eqnarray}
	The integration contour for $d^3 q$ can, for the space of holomorphic
	functions, always be brought to along the real axes, so that the complex
	measure becomes an integration over imaginary $q_o$.  
	The field operator expansion thus contains both the exponential decaying
	and run-away modes, and appears at first sight to cause difficulties
	with large time limit ($t \rightarrow \pm \infty$) as well as with 
	time translation invariance of the operator product matrix 
	elements.  Our study shows the remarkable property, however, that the 
	run-away modes decouple from the physical matrix elements and only the 
	{\em transient} exponential decaying modes remain.  Furthermore, 
	space-time translation invariance is preserved.
	
	This comes about through the two nilpotent vacua in this theory,
	which can be shown to be the $in$ and $out$ vacua of the scattering matrix.  
\\

{\large \bf Physical vacuum matrix elements}

	The physical matrix element of the equal time anti-commutator 
	taken between the $in$ and $out$ vacua remains a simple delta function, 
\begin{equation}
	(\Phi_{out} | \; \{ \psi_{_{L}} (\vec{x}, t), \; \psi^{\dagger}_{_{L}}
				(\vec{y}, t ) \;\} \; | \Phi_{in}) = \delta(\vec{x} - \vec{y})
\end{equation}
	while the corresponding vacuum expectation value of the $\psi_{_{L}}$ 
	field satisfies the causality condition 
\begin{equation}
	( \Phi_{out} | \; \{ \psi_{_{L}} (x) , \bar{\psi}_{_{L}} (y) 
		\} \; |\Phi_{in}) \;= \;  0  
		\;\;\;\;{\rm for} \; (x-y)^2 \;{\rm spacelike}
				\label{eq-causal}
\end{equation}
	As a result of this causal property, the S-matrix operator continues
	to be given by the time-ordered evolutionary operator even in the 
	presence of propagation of tachyonic neutrinos.  

	The time-ordered Green function for the neutrino propagation from 
	space-like separated $x$ to a later $y$, ($y_o > x_o$), 
	when viewed in a different frame with $x^{\prime}_o$ later than
	$y^{\prime}_o$, is according to eq.(\ref{eq-causal}) actually the 
	time-ordered Green function for the antineutrino propagation 
	from $y^{\prime}$ to $x^{\prime}$. 
	The physical matrix elements are correctly described by retarded
	Green functions, so that there are no problems arising from
	the acausal propagation of neutrino backwards in time.
\\ 

{\large \bf Nilpotency of the physical vacua}

	The nilpotency results from the Majorana coupling between the 
	physical $\psi_{_{L}}$ field and the sterile negative metric 
	$\psi^{\prime}_{_{L}}$ field.  This ghost field does not 
	couple to the strong, electromagnetic or weak interactions.  
	It does not occur in the physical S-matrix elements.
	Its only function is to condense with the physical $\psi_{_{L}}$ 
	field to produce an ether ( the $in$ and $out$ vacua ) that 
	support quasi-particle modes with $ v > c $ as well as the 
	transient modes with $k < m$.  

	In this picture, the vacuum is itself a neutrino condensate
	so that in a sense the neutrino's role in our universe 
	transcends that of the photon.  While light illuminates the 
	universe, and is the dominant signal carrier from one part 
	of the universe to another, the low energy neutrino is the 
	weak and largely silent and transparent {\em superluminal} 
	courier that pervades throughout the universe.

	This model has interesting implications for the phenomenology 
	of neutrino oscillation.  In the one flavor case, the physical
	neutrino state (with $p > m$) created at time $t= 0$ propagates
	with no neutrino deficit.  But in a multiflavor generalization,
	the physical neutrino states oscillate as a result of the mixing
	between the flavor fields and the mass eigenstates.  The
	resulting neutrino oscillation phenomology is indistinguishable
	from the usual Dirac flavor mixing oscillation.

\section{Majorana Equation}
\label{sec-Majorana equation}

	Consider the following coupled set of Majorana field equations 
	for the mutually anticommuting left-handed fermion field operators, 
	$\psi_{L}(x,t)$, $\psi_{L}^{\prime}(x,t)$:
	
\begin{eqnarray}
	(\vec{\gamma} \cdot \vec{\nabla} -  \gamma_o \, 
		\frac{\partial}{\partial t}) 
		\psi_{_{L}}( x,t)
			&=&  - m \, \gamma_2 \cdot \psi_{_{L}}^{\prime *}(x,t)
						\label{eq-Majorana-1} \\
	(\vec{\gamma} \cdot \vec{\nabla} -  \gamma_o \, 
		\frac{\partial}{\partial t}) 
		\psi_{_{L}}^{\prime}( x,t)
			&=&   + m \, \gamma_2 \cdot \psi_{_{L}}^{*}(x,t)
						\label{eq-Majorana-2}
\end{eqnarray}
\noindent	where we have taken the usual charge conjugation matrix $C$
	to be $\gamma_2 \, \gamma_4$ and the $\gamma_{\mu}$ matrices are 
	all hermitian, with $ \mu=1, \ldots , 4$ and $\gamma_4 \equiv - i 
	\gamma_o$.   

	From this coupled set of Majorana equations, it is easy to show 
	directly that the fields have {\em space-like} mass-squared.   
	Note that if the sign in eq.(\ref{eq-Majorana-2}) had been 
	opposite, the coupled set of Majorana equations would have 
	led to a normal {\em time-like } mass-squared.

	The equations arise in the Lagrangian
\begin{equation}
	L = - \bar{\psi}_{_{L}} \gamma \cdot \partial \psi_{_{L}}
		+ \bar{\psi^{\prime}}_{_{L}} \gamma \cdot \partial
		\psi^{\prime}_{_{L}} - m \cdot \bar{\psi}_{_{L}}
		C \bar{\psi_{_{L}}^{\prime}}^{T} 
		- m \cdot \psi_{_{L}}^{\prime \, T}  C  \psi_{_{L}}
					\label{eq-Lagrangian}
\end{equation}
 	As far as the Standard Model weak interactions
	are concerned, it is the $\psi_{_{L}}$ flavor field that 
	participates in the charged-current and neutral-current 
	interactions.  The $\psi_{_{L}}^{\prime}$ field does not.  It 
	is a sterile ghost field. 

	As we shall show below, this negative metric of the 
	$\psi^{\prime}$ does not affect the unitarity of the 
	scattering matrix element between the physical $in$ and 
	$out$ vacua.  The sterile ghost modes condense with the 
	$\psi_{_{L}}$ modes to form the nilpotent physical $in$ 
	and $out$ vacua.

\section{Solving equations of motion}
\label{sec-Solving eq motion}

	We may solve the field theory of this coupled set of Majorana 
	field equations by using the usual expansions for the field 
	operators in terms of the time-dependent Heisenberg annihilation 
	and creation operators

\begin{eqnarray}
	\psiL(x,t) &=& \frac{1}{\sqrt{V}} \sum_k \left\{ a_{k,L} 
			\left( {\begin{array}{c}
				 \chi_{_{\ell}} \\ 0 
			   \end{array}} \right)
			\;+\; \bdagkR \left( {\begin{array}{c}
				 \chi_{_{r}} \\ 0 
			   \end{array}} \right)
			 \right\} \, e^{i \vec{k} \cdot \vec{x}}
						\label{eq-psi-Majorana} \\
	\psiL^{\prime}(x,t) &=& \frac{1}{\sqrt{V}} \sum_k \left\{ 
			a^{\prime}_{_{k,L}} 
			\left( {\begin{array}{c}
				 \chi_{_{\ell}} \\ 0 
			   \end{array}} \right)
			\;+\; b^{\prime \dagger}_{_{-k,R}} \; \left( 
				{\begin{array}{c}
				 \chi_{_{r}} \\ 0 
			   \end{array}} \right)
			 \right\} \, e^{i \vec{k} \cdot \vec{x}}
						\label{eq-psi-prime-Majorana} 
\end{eqnarray}

\noindent 	where $\chi_{_{\ell,r}}$ are two-component helicity spinors  
		satisfying the relation

\begin{equation}
	(\pm \, \vec{\sigma} \cdot \vec{k} + k )\chi_{_{\ell, r}} = 0
\end{equation}
\noindent	with the properties
\begin{eqnarray}
		\chi_{_{\ell}} ( - \vec{k} )  &=&  + \, \eta_k^{*}
							\chi_{_{r}} ( \vec{k}) \\
		\chi_{_{r}} ( - \vec{k} )  &=&  - \, \eta_k
							\chi_{_{\ell}} ( \vec{k} ) \\
		i \sigma_2 \,\chi^{*}_{\ell} (-\vec{k} ) &=& - \, \eta_k 
						\chi_{_{\ell}} (\vec{k} ) \\
		i \sigma_2 \,\chi^{*}_{r} (-\vec{k} ) &=& - \, \eta_k^{*} 
						\chi_{_{r}} (\vec{k} )
\end{eqnarray}
\noindent	and $\eta_k$ is the usual phase factor with the property
		that it is odd under parity $\vec{k} \rightarrow - \vec{k}$.
\begin{equation}
		\eta_{k} = \frac{k_x + i k_y}{\sqrt{k^2 - k_z^2}}=- \eta_{-k} 
\end{equation}

	The field equations (eq.(\ref{eq-Majorana-1},\ref{eq-Majorana-2})) 
	lead to equations of motion for the Heisenberg 
	operators:\footnote{hereinafter 
	for brevity of notation we shall write $a \equiv \akL, 
	\; a^{\prime} \equiv a^{\prime}_{_{-k,L}},
	\; b \equiv b_{_{-k,R}}, \; b^{\prime} \equiv 
	b^{\prime}_{_{k,R}}, \eta \equiv \eta_k$ }
\begin{eqnarray}
	- i\; \dot{a} \;\;&=& - k \; a \;\; + \; m \; \eta \; 
				a^{\prime \dagger} \label{eq-adot}\\
	- i\; \dot{a^{\prime}} \;\,&=& - k \; a^{\prime} 
			\; + \; m \; \eta \; a^{\dagger} 
				\label{eq-aprimedot}\\
	- i\; \dot{b^{\dagger}} \,&=& + k \; b^{\dagger} 
			\; + \; m \; \eta^{*} \; b^{\prime} 
				\label{eq-bdagdot}\\
	- i\; \dot{b^{\prime \, \dagger}} &=& + k \; b^{\prime \, \dagger} 
			\; + \; m \; \eta^{*} \; b  \label{eq-bprimedagdot}
\end{eqnarray}

\noindent	The solutions to these equations depend on whether the 
	momentum, $k$, is greater or less than $m$.  For convenience 
	of notation, we shall reserve $p$ for momenta greater than $m$, 
	and $q$ for momenta less than $m$.  
	For $p > m$, we have ($ \omega \equiv \sqrt{p^2 - m^2}$ )
\begin{eqnarray}
	a_{_{p,L}}(t) \;\;&=&  	\sqrt{\frac{p + \omega}{2\omega}} 
			A_{_{p,L}} {\rm e}^{-i \omega t}\;\;\;\;+\; 
				\sqrt{\frac{p - \omega}{2\omega}} \eta_p 
				A^{\prime \dagger}_{_{-p,L}} 
				{\rm e}^{+ i \omega t} \label{a(t)}\\
	a^{\prime}_{_{-p,L}}(t) &=&  	\sqrt{\frac{p + \omega}{2\omega}} 
				A^{\prime}_{_{-p,L}} 
				{\rm e}^{- i \omega t}\;\;+\;\; 
				\sqrt{\frac{p - \omega}{2\omega}} \eta_p 
				A^{\dagger}_{_{p,L}} 
				{\rm e}^{+ i \omega t} \label{aprime(t)}\\
	b^{\dagger}_{_{-p,R}}(t) &=&  	\sqrt{\frac{p + \omega}
				{2\omega}} 
				B^{\dagger}_{_{-p,R}} {\rm e}^{+i \omega t}
				\;\;-\; 
				\sqrt{\frac{p - \omega}{2\omega}} \eta^{*}_p 
				B^{\prime}_{_{p,R}} 
				{\rm e}^{- i \omega t} \label{bdagger(t)}\\
	b^{\prime}_{_{p,R}}(t) \;\;&=&  	\sqrt{\frac{p + \omega}
				{2\omega}} 
				B^{\prime}_{_{p,R}} 
				{\rm e}^{-i \omega t}  \;\;\;-\; 
				\sqrt{\frac{p - \omega}{2\omega}} \eta_p 
				B^{\dagger}_{_{-p,R}} 
				{\rm e}^{+ i \omega t} \label{bprimedagger(t)}
\end{eqnarray}
\noindent	while, for $q < m$, we have the complex unstable modes ($ q \equiv m 
	\cos{\chi}, \;\kappa \equiv \sqrt{m^2 - q^2} = m 
	\sin{\chi}$)
\begin{eqnarray}
	a_{_{q,L}}(t) \;\;&=&  	\frac{1}{\sqrt{2i\sin{\chi}}} 
			\left( 
				{\rm e}^{+ i \chi/2} A_{_{q,L}} 
				{\rm e}^{+ \kappa t} \;\;\;\;\;+\; 
				{\rm e}^{-i \chi/2} \eta_{q} 
					A^{\prime \dagger}_{_{-q,L}} 
				{\rm e}^{- \kappa t} \right) 
			\label{complex-mode1}\\
	a^{\prime}_{_{-q,L}}(t) &=&  	\frac{\sqrt{i}}
			{\sqrt{2\sin{\chi}}} 
			\left(  
				{\rm e}^{+ i \chi/2} \eta_q 
				A^{\dagger}_{_{q,L}} 
				{\rm e}^{+ \kappa t} \;\;\, + \;
				{\rm e}^{-i \chi/2} A^{\prime }_{_{-q,L}}
				{\rm e}^{- \kappa t}\;\;\;\;\right) 
							\label{complex-mode2}\\
	b^{\dagger}_{_{-q,R}}(t) &=&  	\frac{\sqrt{i}}
			{\sqrt{2\sin{\chi}}} 
				\left( {\rm e}^{- i \chi/2} \,
				B^{\dagger}_{_{-q,R}} 
				{\rm e}^{+ \kappa t} \,\;\;\;-\;\; 
				{\rm e}^{+i \chi/2} \eta^{*}_q 
				B^{\prime}_{_{q,R}}
				{\rm e}^{- \kappa t}\right) 
			\label{complex-mode3}\\
	b^{\prime}_{_{q,R}}(t) \;\;&=&  	\frac{\sqrt{i}}
			{\sqrt{2\sin{\chi}}} 
			\left(  
				{\rm e}^{-i \chi/2} B^{\prime }_{_{q,R}} 
				{\rm e}^{- \kappa t}\;-\;\;
				{\rm e}^{+ i \chi/2} \eta_q 
				B^{\dagger}_{_{-q,R}} 
				{\rm e}^{+ \kappa t}
				\;\;\;\right)  
							\label{complex-mode4}
\end{eqnarray}
\noindent	where $A, A^{\prime}, B, B^{\prime}$ are {\it 
	time-independent} Schr$\ddot{o}$dinger operators.

\section{Hamiltonian}
\label{sec-Hamiltonian}

	In terms of the time-dependent Heisenberg operators $a, 
	a^{\prime}, b, b^{\prime}$, the Hamiltonian takes the form
\begin{eqnarray}
	H_o &=& \;\;\;\;\;\;\; \sum_k \; k \; \left( a^{\dagger} a \;-\; 
			a^{\prime\dagger}a^{\prime} 
			\;+\; b^{\dagger} b \;-\; b^{\prime\dagger} 
			b^{\prime} \right) \\
	H_1 &=& - 	m \sum_k \left( a^{\dagger} \;\eta\; 
			a^{\prime\dagger}
			\;+\; a^{\prime} \;\eta^{*}\; a
			\;-\; b^{\dagger} \;\eta\; b^{\prime\dagger}
			\;-\; b^{\prime} \;\eta^{*}\; b
			\right)
\end{eqnarray}
\noindent and we may directly verify that on account of the equations 
	of motion (eq.(\ref{eq-adot}-\ref{eq-bprimedagdot})) we have
\begin{equation}
	\frac{d}{dt} H = 0	\label{eq-dHdt}
\end{equation}

	In terms of the time-independent Schr$\ddot{o}$dinger operators, 
	$A, A^{\prime}, B, B^{\prime}$, the normal-ordered Hamiltonian takes the 
	diagonal form
\begin{eqnarray}
	H	&\equiv&    H_o + H_1  - < \Phi_{in} | \;H\; | \Phi_{in} >  \nonumber \\
		&=&		\;\;\;\sum_{p>m} \left( \;\;\; \omega 
			A_{_{p, L}}^{\dagger} 
				A_{_{p, L}} 	\;\;\;\;-\;
			\omega A_{_{-p, L}}^{\prime \dagger} 
			A_{_{-p, L}}^{\prime} 
			\right.\nonumber  \\
		&&	\left. \;\;\;\;\;\;\;\;\;\;\; + \;\omega 
			B_{_{-p, R}}^{\dagger} B_{_{-p, R}} 
			\;- \;\omega B_{_{p, R}}^{\prime\dagger} 
			B_{_{p, R}}^{\prime}   
			 \;\right) \nonumber \\
		&&	+ \sum_{q < m} \left( -  \kappa 
				A^{\dagger}_{_{q, L}} \eta_q 
				A^{\prime \dagger}_{_{-q, L}} 
				-  \kappa A^{\prime}_{_{-q, L}} 
				\eta^{*}_q A_{_{q, L}}  
				\;+ \;\;q \right.
				 \nonumber \\
		&&	\left. \;\;\;\;\;\;\;\; \;\;\;\; +  \kappa 
				B^{\dagger}_{_{-q, R}} \eta_q 
				B^{\prime \dagger}_{_{q, R}} 
				+ \kappa B^{\prime}_{_{q, R}} 
				\eta^{*}_q B_{_{-q, R}} 
				\,+ \;q \;\right)     
			\label{physical-hamiltonian}
\end{eqnarray}

\section{Quasi-Particle Anti-Commutation Rules}
\label{sec-statistics}

	The Heisenberg operators satisfy the expected equal-time 
	anti-commutation rules for all $\vec{k}, \vec{k'}$
\begin{equation}
	\left. {\begin{array} {lcrcc}
	\{a_{_{k,L}} , a_{_{k', L}}  \} &=& \;\{a^{\prime}_{_{-k,L}}, 
		a^{\prime}_{_{-k',L}} \} &=& 0  \\
	\{a_{_{k,L}} , a^{\prime}_{_{-k', L}}  \} &=& \{a_{_{k,L}}, 
		a^{\prime \dagger}_{_{-k',L}} \} &=& 0 \\
	\{a_{_{k,L}} ,a^{\dagger}_{_{k', L}} \} &=& - \{
		a^{\prime}_{_{k,L}} 
		,a^{\prime \dagger}_{_{-k', L}} \} &=& 
		\delta_{\vec{k}, \vec{k'}} 
	        \end{array}} \right\}
\end{equation}
\noindent and likewise for the anti-particle operators.
	
	Expressed in terms of the quasi-particle operators, however, the 
	anti-commutation rules are different depending on whether $k$ 
	is greater than or less than $m$.  For $p > m$ we have\footnote{
	The anti-commutation rules for the anti-particle $B$ and $B^{\prime}$ 
	operators may be obtained by replacing everywhere $A_{_{k,L}}$ and 
	$A^{\prime}_{_{-k,L}}$ by $B_{_{k,R}}$ and $B^{\prime}_{_{-k,R}}$ 
	respectively.}   
\begin{equation}
	\left. {\begin{array}{lcrcc}
	\{A_{_{p, L}} ,A^{\dagger}_{_{p', L}} \} &=& - \{
		A^{\prime}_{_{p, L}} ,A^{\prime \dagger}_{_{-p', L}} \} 
		&=& 
		\delta_{\vec{p}, \vec{p'}}    \\
	\{A_{_{p, L}} , A_{_{p', L}}  \} &=& \;\;\{A^{\prime}_{_{-p,L}}, 
		A^{\prime}_{_{-p', L}} \} &=& 0  \\
	\{A_{_{p, L}} , A^{\prime}_{_{-p', L}}  \} &=& \{A_{_{p, L}}, 
		A^{\prime \dagger}_{_{-p', L}} \} &=& 0 
		\end{array}} \right\}
\end{equation}
\noindent while for $q < m$, the quasi-particle operators obey the 
	unusual nilpotent relations
\begin{equation}
	\left. {\begin{array} {lcrcc}
	\{A_{_{q, L}} ,A^{\dagger}_{_{q', L}} \} &=& \{
		A^{\prime}_{_{-q, L}} , A^{\prime \dagger}_{_{-q', L}}
		\} &=& 0   \\
	\{A_{_{q, L}} , A_{_{q', L}}  \} &=& \{
		A^{\prime}_{_{-q, L}}, 
		A^{\prime}_{_{-q', L}} \} &=& 0  \\
	\eta^{*}_q \, \{A_{_{q, L}} , A^{\prime}_{_{-q', L}} \} 
		&=& 
		- \eta_q \, \{A^{\dagger}_{_{q, L}}, 
		A^{\prime \dagger}_{_{-q', L}} \} &=& - i  
		\delta_{\vec{q},\vec{q'}}  
		\end{array}} \right\}
\end{equation}
	With these commutation rules, we obtain for $p > m$ the usual 
	operator equation of motion for the Schr$\ddot{o}$dinger operators, 
	$A, A^{\prime}$,	
\begin{equation}
	\left.{\begin{array} {lcl}
	\lbrack H , A_{_{p, L}} \;\;\rbrack  	&=&  - \omega \; 
		A_{_{p, L}} \\
	\lbrack H, A^{\prime}_{_{-p, L}} \rbrack &=&  - \omega \; 
		A^{\prime}_{_{-p, L}}
		\end{array}}			\right\} \;(p > m)
\end{equation}
\noindent etc.
	For $q < m$, however, both the $A_{_{q, L}}$ and its 
	conjugate, $A^{\dagger}_{_{q, L}}$, carry the {\it same} 
	imaginary energy 
\begin{equation}
	\left.{\begin{array} {lcl}
	\lbrack H , A_{_{q, L}} \;\;\rbrack &=&  - i \kappa \; 
			A_{_{q, L}}  \\
	\lbrack H , A^{\dagger}_{_{q, L}} \;\;\rbrack &=&  - i \kappa \; 
				A^{\dagger}_{_{q, L}}  \\
	\lbrack H , A^{\prime}_{_{-q, L}} \rbrack &=&  + i \kappa \; 
			A^{\prime}_{_{-q, L}}  \\
	\lbrack H , A^{\prime \dagger}_{_{-q, L}} \rbrack &=&  + i 
			\kappa \; 
			A^{\prime \dagger}_{_{-q, L}}  
	 \end{array}} \;\;\right\}\;(q < m)
\end{equation}
	and the vacuum carries a complex zero-point energy.  This is not 
	a problem however as the $q < m$ vacuum is nilpotent.

\section{Structure of the Vacuum}
\label{sec-vacuum structure}

	Since the Hamiltonian is diagonal in $k$, we may consider the 
	Hilbert space for each momentum component and study the property 
	of the states in each subspace.
	For convenience, we separate the Hamiltonian in 
	eq.(\ref{physical-hamiltonian}) into the two components, $H_p$, 
	superluminal Hamiltonian with $p > m$, and $H_q$ for complex 
	modes with $q < m$.  

	The physical vacuum is then given by a product of the separate 
	vacua for each momentum component $k$, with the further 
	subdivision into the $a$ and $b$ parts, for the particle and 
	antiparticle sectors.
\begin{equation}
	| \Phi_{in} > = \prod_{p > m} | \Phi^{a}_{p} > \otimes \;
		| \Phi^{b}_{p} >	\;\otimes\; \prod_{q < m} \; 
		| \Phi^{a}_{q; in} > \otimes \;| \Phi^{b}_{q; in} >
\end{equation}

	For $p > m$ Hilbert space, in the particle subspace, ${a}$,  the 
	spectrum of orthogonal states may be enumerated:

\centerline{\begin{tabular}{rcc}
	State  & Norm&  Energy \\ \hline
	$ | \Phi^{a}_p > $ & $+1$ & $0$ \\
	$a^{\dagger}_{_{p, L}} \;\;\; | 0 ) = \;\;\;\;\;\;
		A^{\dagger}_{_{p, L}} 
		\;\;\; | \Phi^{a}_p > $ & $+1$ & $\omega$ \\
	$a^{\prime \dagger}_{_{-p, L}} \; | 0 ) = \;\;\;\;\;\;
		A^{\prime \dagger}_{_{-p, L}} 
		\; | \Phi^{a}_p > $ & $- 1$ & $\omega$ \\
	$A^{\prime \dagger}_{_{-p, L}} \;A^{\dagger}_{_{p, L}} 
		\;| \Phi^{a}_p > $  & $-1$ & $2 \omega$
\end{tabular}}

\noindent	where the vacuum ground state is as usual annihilated by the 
	quasi-particle operators, $A, A^{\prime}$.  The states are all 
	mutually orthogonal.  Note that the pair of (massless) degenerate 
	states are also given in terms of the free field operators acting 
	on the free field vacuum, $| 0 )$.

	The complex Hamiltonian, $H_q$, has an unusual spectrum of states.

\centerline{\begin{tabular}{rcc}
	State & Norm&  Energy \\ \hline
	$ | \Phi^{a}_{q \;; in} > $ & $0$ & $( q + i \kappa )$ \\
	$a^{\dagger}_{_{q, L}} \;\;\; | 0 ) = \;\;\;\;\;
		A^{\dagger}_{_{q, L}} 
		\;\;\; | \Phi^{a}_{q \;;in} > $ & $+1$ & $q$ \\
	$\eta a^{\prime \dagger}_{_{-q, L}} \; | 0 ) = \;\;\;\;\;\;
		A_{_{q, L}} 
		\;\;\, | \Phi^{a}_{q \;;in} > $ & $- 1$ & $q$ \\
	$\;| \Phi^{a}_{q \;out} > \;\equiv\; A_{_{q, L}} 
		\;A^{\dagger}_{_{q, L}} \;| \Phi^{a}_{q \;;in} >   $ 
		& $0$ & $( q - i \kappa)$
\end{tabular}}
\noindent	Here, the pair of degenerate states are mutually orthogonal 
	and have vanishing inner product with both the $in$ and $out$ 
	vacua. They are again related to the free field operators acting 
	on the free field vacuum, $|0)$.

	In this spectrum, the pair of complex energy states are nilpotent.  
	They obey the inner product
\begin{equation}
	 < \Phi^{a}_{q \;;out} | \Phi^{a}_{q \;;in} > \;=\; 1
\end{equation}
	They turn out to be the $in$ and $out$ scattering vacuum states. 
	It is noteworthy that, unlike the usual quasi-particle vacuum, the 
	$in$ vacuum is annihilated by {\em both} $A^{\prime}$ and 
	$A^{\prime \dagger}$ operators, while the $out$ vacuum is 
	annihilated by $A$ and $A^{\dagger}$ operators.
\begin{equation}
	\left.{\begin{array}{lclcl}
	A^{\prime}_{_{-q, L}} \;| \Phi^{a}_{q \; ;in} > 
		&=& 
	A^{\prime \dagger}_{_{-q, L}} \;| \Phi^{a}_{q \; ;in} > 
		&=& 0 \\
	A_{_{q, L}} \;\;\;| \Phi^{a}_{q \; ;out} > 
		&=& 
	A^{\dagger}_{_{q, L}} \;\;\;| \Phi^{a}_{q \; ;out} > 
		&=& 0 
	 \end{array} } \; \right\} \;\; q < m
\end{equation}

	The physical interpretation of the complex vacua becomes clear
	when we consider their relation to the interaction picture
	free field vacuum, $| 0 )$.  Let $U( t_2, t_1)$ denote the time 
	evolution operator that takes any interaction picture state vector 
	from time $t_1$ to $t_2$, 
\begin{equation}
	U(t_2, t_1) = {\rm e}^{-i H ( t_2 - t_1 )} 
			{\rm e}^{+ i H_o ( t_2 - t_1 ) }
\end{equation}
	and consider its action on the $q$ subspace of the Hilbert space 
	in the particle sector ${a}$.  We find that the $in$ vacuum is 
	related to the time evolved state from $t=-\infty$ to $t=0$, while
	$out$ vacuum is related to the time evolved state from $t=0$ to 
	$t=\infty$:
\begin{eqnarray}
	| \Phi^{a}_{q \;;in} >  &=& \lim_{T' \rightarrow \infty} 
		\frac{U( 0, - T') 
		\;| 0 ) }{{\rm e}^{+ \kappa T' - i q T'} 
		\; < \Phi^{a}_{q \;;out} | 0 ) }  \\
	| \Phi^{a}_{q \;;out}> &=& \lim_{T \rightarrow \infty} 
		\frac{U^{\dagger}( T, 0) 
		\;| 0 )}{{\rm e}^{+ \kappa T + i q T} 
		\; < \Phi^{a}_{q \;;in} | 0 ) }  \\
\end{eqnarray}
	The scale factors in the denominator are related to the free field 
	vacuum expectation value of the complete time evolution 
	$U(\infty, -\infty)$(See below).  In deriving this result, we have 
	used the completeness relation for the subspace
\begin{eqnarray}
	\left\{ \frac{}{} \;| \Phi^{a}_{q \;;in}> < \Phi^{a}_{q \;;out} |  \right.
		\;\;\;\;\;\;\;\;\;&+&\; \;\;\;\;\;\;\;\;\;\;\;\;\;\;\; 
		| \Phi^{a}_{q \;;out} > <\Phi^{a}_{q \;;in} | \;  
			\nonumber \\
		\;+\;	\frac{i}{2} \eta^{*} \;A_{_{q, L}} \;| 
			\Phi^{a}_{q \;;in}> < \Phi^{a}_{q \;;out} | \; 
			A^{\prime}_{_{-q, L}} 
		&-&   \frac{i}{2} \eta \;\;A^{\prime \dagger}_{_{-q, L}} 
				\;| \Phi^{a}_{q \;;out}> < 
			\Phi^{a}_{q \;;in} | \; A^{\dagger}_{_{q, L}}  
			\nonumber \\
		 \;-\;	\frac{i}{2} \eta \;\;A^{\dagger}_{_{q, L}} 
				\;| \Phi^{a}_{q \;;in}> 
				< \Phi^{a}_{q \;;out} | \; 
				A^{\prime \dagger}_{_{-q, L}} 
		&+& \left. \frac{i}{2} \eta^{*} \;\;A^{\prime}_{_{-q, L}} 
				\;| \Phi^{a}_{q \;;out}> 
				<  \Phi^{a}_{q \;;in} | 
				\; A_{_{q, L}} \;\;\right\} \;\;\;=\;\;\;  
				I^{a}_q  
\end{eqnarray}
	as well as the identities
\begin{eqnarray}
	A^{\prime}_{_{-q, L}} \;| \Phi^{a}_{q \;;out} > &=&  - i 
			\eta 	\;\;\; A^{\dagger}_{_{q, L}}  
			\;| \Phi^{a}_{q \;;in} > \\
	A^{\prime \dagger}_{_{-q, L}} \;| \Phi^{a}_{q \;;out} > &=&  
			- i \eta^{*} \; A_{_{q, L}} \;
			\;| \Phi^{a}_{q \;;in} > 
\end{eqnarray}

	The physical vacuum state is a combined product eigenstate of 
	both $H_{p}$ and $H_{q}$
\begin{eqnarray}
	| \Phi_{in}> &=&	\prod_{p > m} \left\{
				\left(\cosh \theta_p +  \eta_p  
				\sinh \theta_p
				a^{\dagger}_{p, L} a^{\prime\dagger}_{-p, L} 
					\right) 
				\left(\cosh \theta_p - \eta_p
				\sinh \theta_p
				b^{\dagger}_{-p, R} b^{\prime\dagger}_{p, R}  
					\right) 
						\right\} 	\nonumber 	\\
			&&	\prod_{q < m}\left\{
				\frac{i{\rm e}^{-i\chi}}{2 \sin{\chi}}
					\left( 1 \;+\; {\rm e}^{+i \chi}  
				\;\eta_q \; a^{\dagger}_{q, L} 
				a^{\prime\dagger}_{-q, L} \right) 
				\left( 1 \;-\; {\rm e}^{+i\chi} \;\eta_q 
				\; 	b^{\dagger}_{-q, R}
					b^{\prime\dagger}_{q, R}  \right) 
						\right\} | 0 )
					\label{eq-NJL-vac}
\end{eqnarray}
\noindent	where
\begin{equation}
	\cosh \theta_p	=	\sqrt{\frac{p + \omega}{2\omega}}
\end{equation}

	Eq.(\ref{eq-NJL-vac}) shows a Nambu-Jona-Lasinio\cite{NJL}
	type condensation of the Majorana neutrinos in the physical
	vacuum.  There are crucial differences, however.  The pairing here 
	is with $\psi_{_{L}}$ and $\psi^{\prime}_{_{L}}$ fields of the 
	same handedness, as compared with the pairing of quarks and 
	antiquarks in NJL.  Furthermore, the 
	vacuum supports both the tachyonic (superluminal) quasi-particle 
	modes (for $p > m$), as well as the unstable transients $q < m$.

	The physical S-matrix element taken between the complete physical 
	$in$ and $out$ vacua takes the form ($ x_o > y_o $)
\begin{equation}
	< \Phi_{out} | \psi_{_{L}} (\vec{x}, x_o) \bar{\psi}_{_{L}}
		(\vec{y}, y_o) | \Phi_{in} >
		=  	\frac{( 0 | U( \infty, x_o) \psi_{_{L}} (\vec{x}, x_o) 
			\bar{\psi}_{_{L}}	(\vec{y}, y_o) U( y_o, -\infty) |0)}
			{ (0| U(\infty, - \infty) |0)}
\end{equation}
	so that the relation with the interaction picture becomes manifest.

	For the physical S-matrix, taken with respect to the $in$ 
	and $out$ vacua, the non-vanishing matrix elements are of the type 
\begin{equation}
		{\begin{array} {r c r c l}
		< \Phi_{out} | A_{_{p, L}} \;\; A^{\dagger}_{_{p', L}} \; 
				| \Phi_{in} >
		&=& - <\Phi_{in} | A^{\prime}_{_{-p, L}}  
			A^{\prime \dagger}_{_{-p', L}} 
			\; | \Phi_{out} > &=&  + \delta_{p,p'} \;\;\;\;\;\;\;\; 
				(p, p' > m) \\	
		< \Phi_{out} | A^{\prime \dagger}_{_{-q, L}} \; 
			A^{\dagger}_{_{q', L}} \; 
				| \Phi_{in} >
		&=& <\Phi_{in} | A^{\dagger}_{_{q, L}} 
			\;\; A^{\prime \dagger}_{_{-q', L}} 
			\; | \Phi_{out} > &=& \;\;\, i \eta^{*}_{q} 
			\delta_{q, q'}	\;\;\;\; (q, q' < m)
		  \end{array}}  
\end{equation}
\noindent	As a result, in the physical matrix elements of the 
	time-ordered product $T( \psi_{_{L}} (x) \bar{\psi}_{_{L}} (y))$ 
	between $in$ and $out$ vacua, the exponential runaway modes 
	decouple, leaving behind the only transients ${\rm e}^{- \kappa 
	|x_o - y_o| } $.  Co-incidentally, this decoupling of the runaway 
	modes restores the translational invariance of the time-ordered 
	products taken between the $in$ and $out$ states. 

	These features are important outcome of our coupled Majorana field 
	equations.  In addition to the quasi-particle superlumninal modes, 
	the proper quantization of the tachyons reveals the presence of 
	complex transients  ${\rm e}^{ - \kappa |x_o - y_o|}$.  Earlier 
	work in condensed matter physics\footnote{In condensed matter 
	physics, the work of Chiao and others \cite{tachyon-medium} have 
	shown that superluminal propagation occurs in a medium with inverted 
	population.  The inverted population lead to unstable modes that are 
	necessary for causality (see Aharonov and 
	co-workers\cite{tachyon-causal}). }
	suggests that we interpret this transient observed at $x_o$ as being 
	associated with the population inversion at the earlier time $y_o$.

\section{Micro-Causality}
\label{sec-causality}

	The neutrino field, $\psiL$, satisfies the normal causality 
	relations with respect to the S-matrix element between the $in$ 
	and $out$ vacua.
\begin{equation}
	\left. {\begin{array} {l c l}
              < \Phi_{out} | \{ \psiL(x), \bar{\psi}_{_{L}} (y) \} 
			|\Phi_{in} > &=& 0  \\
              <\Phi_{out} | \{ \bar{\psi^{\prime}}_{_{L}}(x), 
			\bar{\psi}_{_{L}} (y) \} 
				|\Phi_{in} > &=& 0 
              \end{array}} \right\}	\;\;\;\;\; (x - y)^2 \; 
			{\rm spacelike}
\end{equation}

	The equal time relations of the physical $\psiL$ field remain valid
	with respect to the S-matrix elements
\begin{equation}
	<\Phi_{out} | \{ \psiL(\vec{x},0), \psi^{\dagger}_{_{L}} (\vec{y},0) 
	\} |\Phi_{in}> \;=\; \delta ( \vec{x} - \vec{y})
\end{equation}

	The time-ordered Green function for the physical $\psiL$ field is
	given by
\begin{equation}
	<\Phi_{out} | T( \psi_{_{L}} (x) \bar{\psi}_{_{L}}(y)) |\Phi_{in}>
	=  \int \frac{d^4 k}{(2\pi)^4} \frac{- \vec{\gamma}\cdot \vec{k}
		- \gamma_o \cdot k_o}{\vec{k}\,{}^2 - k_o^2 - m^2 
	- i\epsilon} \; {\rm e}^{i \vec{k} \cdot (\vec{x} - \vec{y}) 
					- i k_o (x_o - y_o)}
\end{equation}
\noindent	and by closing the contour in the lower half $k_o$ plane for 
	$x_o > y_o£¬$ we recover the quasi-particle contributions from 
	$k > m$, as well as the transient mode contributions from $k < m$.

\section{Time Evolved State}
\label{sec-Time Evolved State}

	In this model, the negative metric $\psi^{\prime}_{_{L}}$ is a 
	sterile field, it is not created by the usual weak and 
	electromagnetic interactions.  Its role is to condense with the 
	physical $\psi_{_{L}}$ field to form the $in$ and $out$ vacua.
	In the physical S-matrix elements that are to be taken between $in$ and 
	$out$ vacua, only $\psi^{\dagger}_{_{L}}$ and $\psi_{_{L}}$ operators are 
	present to create and annihilate physical neutrinos.  

	The physical neutrino and anti-neutrino states with $p > m$ created at $t=0$ from 
	the in-vacuum is represented by\footnote{
		Transient states with $q < m$ are exponentially damped in time after creation
		at $t=0$, and do not contribute to neutrino oscillations.}
\begin{eqnarray}
	| \nu_{L}, \vec{p}, 0 >  &=& 
		\int \frac{ d^3 x}{\sqrt{V}} \; \psi^{\dagger}_{_{L}}
		( \vec{x}, 0 ) \;u_{_{L}}\; | \Phi_{in}>  \;{\rm e}^{ i \vec{p}\cdot
		\vec{x}  } 	\label{eq-neutrino-initialstate}\\
	| \bar{\nu}_{R}, \vec{p}, 0 >  &=& 
		\int \frac{ d^3 x}{\sqrt{V}}  
		\;\tilde{\psi}_{_{L}}( \vec{x}, 0 ) \;C\; v_{_{R}} \;| \Phi_{in}> 
		\;{\rm e}^{ i \vec{p}\cdot \vec{x}  } 
				\label{eq-antineutrino-initialstate}
\end{eqnarray}
	where
\begin{equation}
	\begin{array} {lclllcl}
	u_{_{L}} &=& \left( \begin{array} {c}
				c \chi_{_{\ell}} + s \chi_{_{r}} \\
				0 
				\end{array}\right) &,&
	u_{_{R}} &=& \left( \begin{array} {c}
				0 \\
				c \chi_{_{r}} - s \chi_{_{\ell}}  
				\end{array}  \right)	\\
	v_{_{R}} &=& \left( \begin{array} {c}
				c \chi_{_{r}} - s \chi_{_{\ell}} \\
				0 
				\end{array}\right) &,&
	v_{_{L}} &=& \left( \begin{array} {c}
				0 \\
				c \chi_{_{\ell}} + s \chi_{_{r}}  
				\end{array}\right) 
	\end{array}
\end{equation}
	are the positive and negative energy spinor solutions of the equation
\begin{eqnarray}
	i ( \vec{\gamma} \cdot \vec{p} - \gamma_o \omega ) \; u_{_{L, R}} &=& 
			\; m \gamma_5 \; u_{_{R, L}} \\
	i ( \vec{\gamma} \cdot \vec{p} + \gamma_o \omega ) \; v_{_{L, R}} &=& 
			\; m \gamma_5 \; v_{_{R, L}} \\
\end{eqnarray}
	and we have used the abbreviation $c \equiv \sqrt{\frac{p + 
   	\omega}{2 p}}$ and $ s  \equiv \sqrt{\frac{p - \omega}{2 p}}$, 
	with $c^2 + s^2 = 1$.

\begin{eqnarray}
	| \nu_{L}; \vec{p}; t > &=&  {\rm e}^{- i H t} \; 
			| \nu_{L}; \vec{p}; 0 >  \\
		&=&   {\rm e}^{- i \omega t} \; 
			\left(c \; \cosh \theta_p A^{\dagger}_{_{p, L}} 
			- s \; \sinh \theta_p \eta_p \;B^{\prime \dagger}_{_{p, R}} \right)
						\;| \Phi_{in} >
				\label{eq-neutrino-state}
\end{eqnarray}

	Eq.(\ref{eq-neutrino-state}) shows that {\em in the one-flavor case}, 
	the physical neutrino 
	state is actually a superposition of the physical quasi-particle 
	mode, $A_{_{q,L}}$, and the negative metric quasi-antiparticle 
	mode, $B^{\prime}_{_{q,R}}$.  The time evolution of this state
	does not show an oscillation into a right-handed neutrino state.
	Instead the state behaves as an eigenstate of the Hamiltonian,
	with the usual (superluminal) time dependence.  There is thus
	{\em no neutrino deficit} as it propagates in the ether.
	The state remains of positive unit norm as time evolves.

	In establishing the time evolution of this Schr$\ddot{o}$dinger state 
	vector, we note a useful set of identities with respect to the states 
	created by the fields
\begin{eqnarray}
	\int d^3x\; \bar{\psi}_{_{L}}(\vec{x},t) \;|\Phi_{in} > \;\gamma_5 u_{R} 
		\,{\rm e}^{i p \cdot x}  
		&=& - \int d^3 x \; \tilde{\psi^{\prime}}_{_{L}}(\vec{x}, t)  
			\;|\Phi_{in}> \;C \, u_{L} \,{\rm e}^{i p \cdot x}   \\
	\int d^3x\; \bar{\psi^{\prime}}_{_{L}}(\vec{x}, t) \;|\Phi_{in}>\; \gamma_5 u_{R} 
		\,{\rm e}^{i p \cdot x}  
		&=& - \int d^3 x \; \tilde{\psi}_{_{L}}(\vec{x}, t) 
		 	\;|\Phi_{in}>  \;C \, u_{L} \,{\rm e}^{i p \cdot x} 
\end{eqnarray}

\section{Reduction Formulae}
\label{sec-reduction formulae}

	The scattering matrix element for the physical neutrino field may be
	expressed in terms of the reduction formulae.  Let $| \nu_{L}; p >_{in}$
	denote the incoming scattering state for the superluminal neutrino 
	with space-like momentum $p$, then
\begin{eqnarray}
	| \nu_{L}; p >_{in}  &=&  \lim_{t \rightarrow -\infty}
					\int \frac{d^{3}x }{\sqrt{V}} \; \bar{\psi}_{_{L}} (x)
						\;| \Phi_{in} > \;
				\gamma_4 \; u_{_{L}} {\rm e}^{i p \cdot x} \\
			&=&  \left(c \; \cosh \theta_p A^{\dagger}_{_{p, L}} 
				- s \; \sinh \theta_p \eta_p \;B^{\prime \dagger}_{_{p, R}} \right)
						\;| \Phi_{in} > 
				\label{eq-neutrino-in-state} 
\end{eqnarray}

\section{3-Flavor Phenomenology}
\label{sec-phenomenology}

	The toy model we have considered so far consists of one left-handed
	flavor mixing with a sterile left-handed neutrino field.  The
	toy model may be made realistic by having three flavors of 
	$\psi_{_{L}}^{\alpha} \, , \alpha = ( {\rm e}, \mu, \tau  )$ 
	fields coupled to the sterile left-handed $\psi^{\prime}_{_{L}}$ field:\footnote{
		For simplicity, we have not considered the more general case with three different 
		tachyonic masses.  This may be obtained by coupling the three flavor neutrinos to 
		three sterile neutrinos.
		}
\begin{eqnarray}
	(\vec{\gamma} \cdot \vec{\nabla} -  \gamma_o \, 
		\frac{\partial}{\partial t}) 
		\; \psi_{_{L}}^{\alpha} &=&
		- m \; u^{\alpha}_{3} \; \gamma_{2} \; 
		\psi_{_{L}}^{\prime *} \\
	(\vec{\gamma} \cdot \vec{\nabla} -  \gamma_o \, 
		\frac{\partial}{\partial t}) 
		\; \psi_{_{L}}^{\prime} &=&
		+ m \; u^{3}_{\alpha} \; \gamma_{2} \; \psi_{_{L}}^{\alpha}
\end{eqnarray}
	\noindent Here $u^{\alpha}_{\beta}$ is the unitary mixing matrix
	that rotates from the Standard Model flavor basis to the
	eigenstates of the Majorana equation: two massless left-handed
	neutrino fields, $\psiL^{(1)}, 
	\psiL^{(2)}$, together with the tachyonic massive 4-component 
	field made up of $\psiL^{(3)}$ and the sterile $\psiL^{\prime}$, where
\begin{eqnarray}
	(\vec{\gamma} \cdot \vec{\nabla} -  \gamma_o \, 
		\frac{\partial}{\partial t}) 
		\; \psi_{_{L}}^{(1)} &=&  0 \\
	(\vec{\gamma} \cdot \vec{\nabla} -  \gamma_o \, 
		\frac{\partial}{\partial t}) 
		\; \psi_{_{L}}^{(2)} &=&  0 \\
	(\vec{\gamma} \cdot \vec{\nabla} -  \gamma_o \, 
		\frac{\partial}{\partial t}) 
		\; \psi_{_{L}}^{(3)} &=&
		- m \; \gamma_{2} \; \psi_{_{L}}^{\prime *} \\
	(\vec{\gamma} \cdot \vec{\nabla} -  \gamma_o \, 
		\frac{\partial}{\partial t}) 
		\; \psi_{_{L}}^{\prime} \;\;&=&
		+ m \; \gamma_{2} \; \psi_{_{L}}^{(3)}
\end{eqnarray}
	The time evolution of the neutrino flavor $\alpha$ state now is of the form
\begin{equation}
	| \nu^{\alpha}; p ; t > \;=\; u^{\alpha}_{1} {\rm e}^{-i p t} | \nu^{(1)}; p; 0 >
			\;+\; u^{\alpha}_{2} {\rm e}^{-i p t} | \nu^{(2)}; p; 0 >
			\;+\; u^{\alpha}_{3} {\rm e}^{-i \omega t} | \nu^{(3)}; p; 0 >
\end{equation}
	and the oscillation into the neutrino flavor $\beta$ state is found by looking
	at the overlap 
\begin{eqnarray}
	< \nu^{\beta}; p; 0 | \nu^{\alpha}; p ; t > \;&=&\; ( u^{1}_{\beta} u^{\alpha}_{1} +
		u^{2}_{\beta} u^{\alpha}_{2} ) {\rm e}^{ -ipt} \;+\; u^{3}_{\beta} u^{\alpha}_{3}
		{\rm e}^{-i \omega t}  \\
		&=& {\rm e}^{-ipt} \left( \delta^{\alpha}_{\beta} + 2 i \, u^{3}_{\beta} u^{\alpha}_{3}
		\,{\rm e}^{i (p - \omega) t / 2} \, \sin{ \frac{(p - \omega) t}{2}} \right) \\
		&\approx& {\rm e}^{-ipt} \left( \delta^{\alpha}_{\beta} + 2 i \, u^{3}_{\beta} u^{\alpha}_{3}
		\,{\rm e}^{i m^2 t / 4p} \, \sin{ \frac{m^2 t}{4p}}
		 \;\;\right)	\label{eq-3-flavor-osc}
\end{eqnarray}
	Eq.(\ref{eq-3-flavor-osc}) is indistinguishable from the usual Dirac flavor oscillation formula.  
	Even though the physical neutrino $\nu^{(3)}$ has a tachyonic mass,  the neutrino oscillation rates 
	take the same form as in the temporal mass case
\begin{eqnarray}
	P_{\alpha \rightarrow \alpha} &=& \left(1 - 2 |u^{\alpha}_{3}|^2 \left( \sin{\frac{m^2 t}{4}}\right)^2 \right)^2
			\;+\; |u^{\alpha}_{3}|^4 \left( \sin{\frac{m^2 t}{2}} \right)^2 \\
	P_{\alpha \rightarrow \beta}	&=&  4 |u^{3}_{\beta}|^2 |u^{\alpha}_{3}|^2 \left(\sin{\frac{m^2 t}{4p}} \right)^2
\end{eqnarray}

\section{Conclusion}

	In this paper, we have studied the canonical quantization of an interacting tachyonic majorana field theory.
	We have shown how micro-causality is preserved in the physical matrix elements taken between the 
	$in$ and $out$ states.  In the single flavor case, the physical neutrino state evolves as a tachyonic
	mass state, with no neutrino deficit.  In the 3-flavor case, there can be neutrino oscillation due to
	flavor mixing.  The phenomenology is indistinguishable from the usual timelike oscillations.

	Studies of the effect of the tachyonic mass on the $\beta$-decay spectrum are underway, and will be reported
	in a future publication.

\end{document}